# LIMITATIONS OF MARKOV CHAIN MONTE CARLO ALGORITHMS FOR BAYESIAN INFERENCE OF PHYLOGENY

By Elchanan Mossel[1] and Eric Vigoda[2]

*University of California at Berkeley and Georgia Institute of Technology*

Markov chain Monte Carlo algorithms play a key role in the Bayesian approach to phylogenetic inference. In this paper, we present the first theoretical work analyzing the rate of convergence of several Markov chains widely used in phylogenetic inference. We analyze simple, realistic examples where these Markov chains fail to converge quickly. In particular, the data studied are generated from a pair of trees, under a standard evolutionary model. We prove that many of the popular Markov chains take exponentially long to reach their stationary distribution. Our construction is pertinent since it is well known that phylogenetic trees for genes may differ within a single organism. Our results shed a cautionary light on phylogenetic analysis using Bayesian inference and highlight future directions for potential theoretical work.

**1. Introduction.** Bayesian inference of phylogeny has had a significant impact on evolutionary biology [14]. There is now a large collection of popular algorithms for Bayesian inference, including programs such as MrBayes [13], BAMBE [22] and PAML [21, 25]. All of these algorithms rely on Markov chain Monte Carlo methods to sample from the posterior probability of a tree given the data. In particular, they design a Markov chain whose stationary distribution is the desired posterior distribution, computed using the likelihood and the priors. Hence, the running time of the algorithm depends on the convergence rate of the Markov chain to its stationary distribution.

Therefore, reliable phylogenetic estimates depend on the Markov chains reaching their stationary distribution before the phylogeny is inferred. A

Received October 2005; revised June 2006.
[1]Supported by a Miller fellowship in Computer Science and Statistics, by a Sloan Fellowship in Mathematics and by NSF Grants DMS-05-04245, DMS-05-28488 and DMS-05-48249 (CAREER).
[2]Supported by NSF Grant CCR-0455666.

variety of schemes (such as multiple starting points [12]) and increasingly sophisticated algorithms (such as Metropolis-coupled Markov chain Monte Carlo in MrBayes [13]) are heuristically used to ensure that the chains converge quickly to their stationary distribution. However, there is no theoretical understanding of the circumstances which the Markov chains will converge quickly or slowly. Thus, there is a critical need for theoretical work to guide the multitude of phylogenetic studies using Bayesian inference.

We consider a setting where the data are generated at random, under a standard evolutionary model, from the mixture of two tree topologies. Such a setting is extremely relevant to real-life data sets. A simple example is molecular data consisting of DNA sequences for more than one gene. It is well known that phylogenetic trees can vary between genes (see [11] for an introduction).

We prove that in the above setting, many of the popular Markov chains take extremely long to reach their stationary distribution. In particular, the convergence time is exponentially long in the number of characters of the data set (a character is a sample from the distribution on the pair of trees). This paper appears to be the first theoretical work analyzing the convergence rates of Markov chains for Bayesian inference. Previously, Diaconis and Holmes [6] analyzed a Markov chain whose stationary distribution is uniformly distributed over all trees, which corresponds to the case with no data.

Our work provides a cautionary tale for Bayesian inference of phylogenies and suggests that if the data contains more than one phylogeny, then great caution should be used before reporting the results from Bayesian inference of the phylogeny. Our results clearly identify further theoretical work that would be of great interest. We discuss possible directions in Section 3.

The complicated geometry of "tree space" poses highly nontrivial difficulties in analyzing maximum likelihood methods on phylogenetic trees, even for constant tree sizes.

Initial attempts at studying tree space include work by Chor et al. [3] which constructs several examples where multiple local maxima for likelihood occur. Their examples use nonrandom data sets (i.e., not generated from any model) on a four species taxa and the multiple optima occur on a specific tree topology, differing only in the branch lengths.

A different line of work beginning with Yang [24] analytically determines the maximum likelihood over rooted trees on three species and binary characters. Since then, some sophisticated tools from algebraic geometry have been used to study the likelihood function and other polynomials on tree space (see, e.g., [7, 23]). It appears that the main result on tree spaces needed in this paper does not follow directly from the algebraic geometry methodology.

NOTE 1. The results proved here were presented to a wide scientific audience in a short report published in [20].

1.1. *Formal statement of results for binary model.* We present the formal definitions of the various notions and then precisely state our results.

Let $\Omega$ denote the set of all phylogenetic trees for $n$ species. Combinatorially, $\Omega$ is the set of (unrooted) trees $T = (V, E)$ with internal degree 3 and $n$ leaves.

The *likelihood* of a data set for a tree is defined as the probability that the tree generates the data set, under a given evolutionary model. For simplicity, we first discuss our results for one of the simplest evolutionary models, known as the Cavender–Farris–Neyman (CFN) model [2, 8, 19], which uses a binary alphabet. Our results extend to the Jukes–Cantor model with a four-state alphabet and to many other mutation models, as discussed later.

For a tree $T \in \Omega$, let $V_{\text{ext}}$ denote the leaves, $V_{\text{int}}$ denote the internal vertices, $E$ denote the edge set and $\mathbf{p}\colon E \to [0, 1/2]$ denote the edge mutation probabilities. The data are a collection of binary assignments to the leaves. Under the CFN model, the probability of an assignment $D\colon V_{\text{ext}} \to \{0,1\}$ is

$$\Pr(D|T,\mathbf{p}) = \tfrac{1}{2} \sum_{\substack{D' \in \{0,1\}^V : \\ D'(V_{\text{ext}}) = D(V_{\text{ext}})}} \prod_{\substack{e=(u,v) \in E(T): \\ D'(u)=D'(v)}} (1 - \mathbf{p}(e)) \prod_{\substack{e=(u,v) \in E(T): \\ D'(u) \neq D'(v)}} \mathbf{p}(e).$$

Below, we will further assume that the marginal distribution at any node of the tree is given by the uniform distribution on $\{0,1\}$.

Note that when $\mathbf{p}(e)$ is close to zero, the endpoints are likely to receive the same assignment, whereas when $\mathbf{p}(e)$ is close to $1/2$, the endpoints are likely to receive independently random assignments. Under the "molecular clock" assumption, edge $e$ has length proportional to $-\log_2(1 - 2\mathbf{p}(e))$.

An algorithmic method for generating a character $D$ for a tree $T$ with weights $\mathbf{p}$ is first to generate a uniformly random assignment for an arbitrary vertex $v$. Then beginning at $v$, for each edge $e = (v, w)$, given the assignment to one of the endpoints, the other endpoint receives the same assignment with probability $1 - \mathbf{p}(e)$ and a different assignment with probability $\mathbf{p}(e)$.

Finally, for a collection of data $\mathbf{D} = (D_1, \ldots, D_N)$,

$$\Pr(\mathbf{D}|T,\mathbf{p}) = \prod_{D \in \mathbf{D}} \Pr(D|T,\mathbf{p})$$

$$= \exp\!\left(\sum_{D \in \mathbf{D}} \log(\Pr(D|T,\mathbf{p}))\right).$$

Now, applying Bayes rule, we can write the posterior probability of a tree given the data,

$$\Pr(T|\mathbf{D}) = \frac{\int_{\mathbf{p}} \Pr(\mathbf{D}|T,\mathbf{p}) \Psi(T,\mathbf{p})\, d\mathbf{p}}{\Pr(\mathbf{D})}$$

$$= \frac{\int_{\mathbf{p}} \Pr(\mathbf{D}|T,\mathbf{p})\Psi(T,\mathbf{p})\,d\mathbf{p}}{\sum_{T'} \int_{\mathbf{p}} \Pr(\mathbf{D}|T',\mathbf{p})\Psi(T',\mathbf{p})\,d\mathbf{p}},$$

where $\Psi(T,\mathbf{p})$ is the prior density on the space of trees, so that

$$\sum_T \int_{\mathbf{p}} \Psi(T,\mathbf{p})\,d\mathbf{p} = 1.$$

Since the denominator is difficult to compute, Markov chain Monte Carlo is used to sample from the above distribution. For an introduction to Markov chains in phylogeny, see [9].

The algorithms for Bayesian inference differ in their choice of Markov chain to sample from the distribution and in their choice of prior. In practice, the choice of an appropriate prior is an important concern. Felsenstein [9] gives an introduction to many of the possible priors. Rannala and Yang [21] introduce a prior based on a birth–death process, whereas Huelsenbeck and Ronquist's program MrBayes [13] allows the user to input a prior (using either uniform or exponential distributions). Our results hold for all these popular priors and only require that the priors are so-called $\varepsilon$-regular for some $\varepsilon > 0$, in the sense that

$$\text{for all } T \text{ and } \mathbf{p} \quad \Psi(T,\mathbf{p}) \geq \varepsilon.$$

Each tree $T \in \Omega$ is given a *weight*,

$$w(T) = \int_{\mathbf{p}} \Pr(\mathbf{D}|T,\mathbf{p})\Psi(T,\mathbf{p})\,d\mathbf{p}.$$

Computing the weight of a tree can be done efficiently via dynamic programming in cases where $\Psi$ admits a simple formula. In other cases, numerical integration is needed. See [9] for background.

The transitions of the Markov chain $(T_t)$ are defined as follows. From a tree $T_t \in \Omega$ at time $t$:

1. Choose a neighboring tree $T'$. See below for design choices for this step.
2. Set $T_{t+1} = T_t$ with probability $\min\{1, w(T')/w(T)\}$, otherwise set $T_{t+1} = T_t$.

This is an example of the standard Metropolis algorithm. The acceptance probability of $\min\{1, w(T')/w(T)\}$ implies that the stationary distribution $\pi$ is proportional to the weights $w$, that is, for $T \in \Omega$,

$$\pi(T) = \frac{w(T)}{\sum_{T' \in \Omega} w(T')}.$$

Three natural approaches for connecting the tree space $\Omega$ are *nearest-neighbor interchanges* (NNI), *subtree pruning and regrafting* (SPR) and *Tree-Bisection-Reconnection* (TBR). In NNI, one of the $n-3$ internal edges is chosen at random and the four subtrees are reconnected randomly in one of the three ways; see Figure 1 for an illustration. In SPR, a random edge is chosen, one of the two subtrees attached to it is removed at random and reinserted along a random edge in the remaining subtree; see Figure 2. In TBR, one of the edges of the tree is removed to obtain two trees. Then the two trees are joined by an edge connecting two midpoints of edges of the two trees. The SPR moves are a subset of the TBR moves, but the two move sets are identical when the tree has fewer than six leaves. We refer to the above chains as Markov chains with discrete state space and NNI, SPR, TBR transitions, respectively.

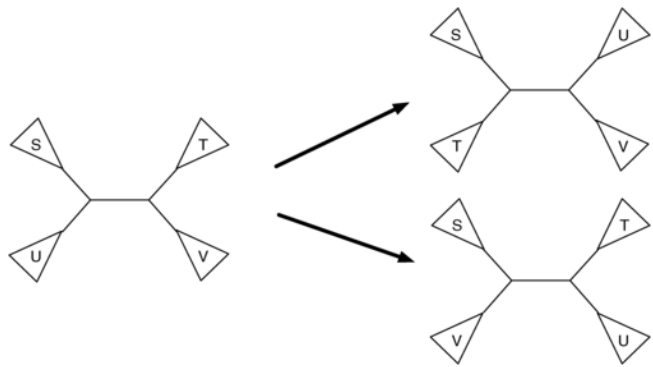

Fig. 1. *Illustration of NNI transition. An internal edge has four subtrees attached. The transition reattaches the subtrees randomly. Since the trees are unrooted, there are three ways of attaching the subtrees, one of which is the same as the original tree.*

Some Markov chains instead walk on the continuous state space where a state consists of a tree with an assignment of edge probabilities. Our results extend to chains with continuous state space where transitions only modify the tree topology by an NNI, SPR or TBR transition and edge probabilities

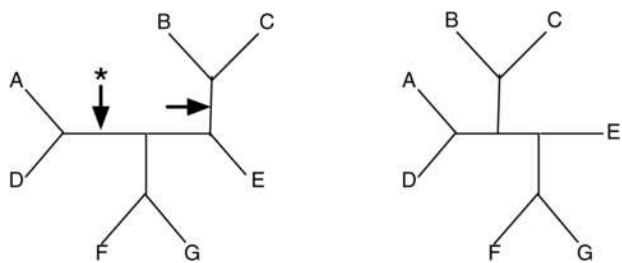

Fig. 2. *Illustration of SPR transition. The randomly chosen edge is marked by an arrow. The subtree containing B and C is removed and reattached at the random edge marked by a starred arrow. The resulting tree is illustrated.*

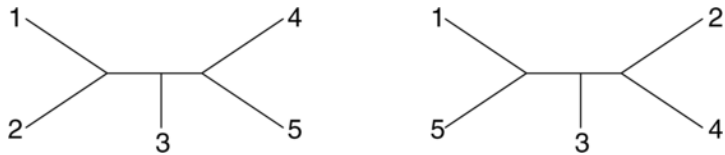

Fig. 3. *The trees $T_1$ and $T_2$.*

are always in $(0, 1/2)$. Some examples of continuous state space chains are [17] and [4, 16].

The *mixing time* of the Markov chain $T_{\text{mix}}$ is defined to be the number of transitions until the chain (from the worst initial state) is within total variation distance $\leq 1/4$ from the stationary distribution.

We consider data coming from a mixture of two trees, $T_1(a, a^2)$ and $T_2(a, a^2)$. $T_1$ is given by $((12), 3), (45)$, while $T_2$ is given by $((15), 3), (24)$; see Figure 3. On the trees $T_1(a, a^2)$ and $T_2(a, a^2)$, we have two edge probabilities, one for those edges incident to the leaves and a different one for internal edges. We let the probability of edges going to the leaves be $a^2$ and let the internal edges have probability $a$, where $a$ will be chosen to be a sufficiently small constant. The trees $T_1(a, a^2)$ and $T_2(a, a^2)$ will have small edge probabilities, as commonly occurs in practice.

We let $\mathcal{D}_1$ be the distribution of the data according to $T_1(a, a^2)$ and $\mathcal{D}_2$ the distribution according to $T_2(a, a^2)$. We let $\mathcal{D} = 0.5(\mathcal{D}_1 + \mathcal{D}_2)$ and consider a data set consisting of $N$ characters.

We prove the following theorem:

THEOREM 1. *For all sufficiently small $a > 0$, there exists a constant $c > 0$ such that for all $\varepsilon > 0$, the following holds. Consider a data set with $N$ characters, that is, $\mathbf{D} = (D_1, \ldots, D_N)$, chosen independently from the distribution $\mathcal{D}$. Consider the Markov chains on tree topologies defined by nearest-neighbor interchanges or subtree pruning and regrafting. Then with probability at least $1 - \exp(-cN)$ over the data generated, the mixing time of the Markov chains, with priors which are $\varepsilon$-regular, satisfies*

$$T_{\text{mix}} \geq c\varepsilon \exp(cN).$$

Note that $\varepsilon$ only has a small effect on this lower bound for the mixing time.

REMARK 2. Our results show that the mixing times of the Markov chains are exponentially slow when the data are generated by a mixture of two trees and there are $n = 5$ leaves. One would expect that the same phenomenon holds for a much more general class of mixture distributions and for trees with larger numbers of leaves, but except for some trivial extensions of our results, it is generally unknown when this is the case. This suggests the following open problems:

- Our proof relies on proving that the likelihood function, under the mixture distribution $\mathcal{D}$, obtains values on trees $T_1$ and $T_2$ that are larger than the maximal value on any other tree. This raises the following question. For $n = 5$, for what branch lengths (i.e., edge probabilities) in the generating trees $T_1$ and $T_2$, under the resulting mixture distribution, is the likelihood at any tree whose topology is different from that of $T_1$ and $T_2$ smaller than the maximal value obtained at $T_1$ and at $T_2$? In such cases, the mixing time will be exponential in the number of characters.
- More challenging is the case when there are significantly more than five leaves. For example, we do not know the answer to the following problem. Suppose that the data are generated from a mixture of trees $T_1'$ and $T_2'$ on $n > 5$ leaves and that there exists a subset $S$ of five leaves, where the induced subtree of $T_1'$ on $S$ is $T_1(a, a^2)$ and the induced subtree of $T_2'$ on $S$ is $T_2(a, a^2)$. Does this imply that the mixing time is exponential in the sequence length (i.e., number of characters)?

1.2. *General mutation models.* As mentioned above, our theorem is valid for many of the mutation models discussed in the literature. We now define these models and derive some elementary features of them that will be used below. In the general case, it is easier to define the evolution model on rooted trees. However, since we will only discuss reversible models, the trees may be rooted arbitrarily. Moreover, for general models, we consider rooted trees with edge lengths, as opposed to unrooted trees with edge probabilities.

The mutation models are defined on a finite character set $\mathcal{A}$ of size $q$. We will denote the letters of this alphabet by $\alpha, \beta$, and so on. The mutation model is given by a $q \times q$ *mutation rate matrix* $Q$ that is common to all edges of the tree, along with a vector $(l(e))_{e \in E(T)}$ of edge lengths. The mutation along edge $e$ is given by

$$\exp(l(e)Q) = I + l(e)Q + \frac{l^2(e)Q^2}{2!} + \frac{l^3(e)Q^3}{3!} + \cdots.$$

Thus, the probability of an assignment $D: V_{\text{ext}} \to \mathcal{A}$ is

$$\Pr(D|T, l) = \sum_{\substack{D' \in \mathcal{A}^V: \\ D'(V_{\text{ext}}) = D(V_{\text{ext}})}} \pi_{D'(r)} \prod_{e=(u,v) \in E(T)} [\exp(l(e)Q)]_{D'(u), D'(v)},$$

where all the edges $(u, v)$ are assumed to be directed away from the root $r$ and where $\pi$ denotes the initial distribution at the root.

It is well known that the CFN model is a special case of the model above where $Q = \begin{pmatrix} -1 & 1 \\ 1 & -1 \end{pmatrix}$ and $\mathbf{p}(e) = (1 - \exp(-l(e)))/2$.

We will further make the following assumptions:

ASSUMPTION 1. 1. The Markov semigroup $(\exp(tQ))_{t\geq 0}$ has a unique stationary distribution given by $\pi$ such that $\pi_\alpha > 0$ for all $\alpha$. Moreover, the semigroup is *reversible* with respect to $\pi$, that is, $\pi_\alpha Q_{\alpha,\beta} = \pi_\beta Q_{\beta,\alpha}$ for all $\alpha$ and $\beta$.

2. The character at the root has marginal distribution $\pi$. By the reversibility of the semigroup, it then follows that the marginal distribution at every node is $\pi$.

3. The rate of transitions from a state is the same for all states. More formally, there exists a number $q$ such that for all $\alpha$,

$$\sum_{\beta \neq \alpha} Q_{\alpha,\beta} = -Q_{\alpha,\alpha}. \tag{1}$$

In fact, by rescaling the edge length of all edges, we may assume without loss of generality that $Q_{\alpha,\alpha} = 1$ for all $\alpha$.

REMARK 3. Parts 1 and 2 of the assumption together imply that we obtain the same model for all possible rootings of any specific tree. Thus, the model is, in fact, defined on unrooted trees.

REMARK 4. It is straightforward to check that our assumptions include as special cases the CFN model, the Jukes–Cantor model, Kimura's two-parameter model and many other models. See [18] for an introduction to the various evolutionary models.

1.3. *Statement of the general theorem.*

DEFINITION 5. Let $\mathcal{T}$ be the space of all trees and edge lengths on five leaves. We say that a prior density $\Psi$ on $\mathcal{T}$ is $(\varepsilon, a)$-*regular* if for every $T$ and $\mathbf{l}$ where $\mathbf{l}(e) \leq 2a$ for all $e$, it holds that $\Psi(T, \mathbf{l}) \geq \varepsilon$.

REMARK 6. For all $a > 0$, all of the priors used in the literature are $(a, \varepsilon)$-regular for an appropriate value of $\varepsilon = \varepsilon(a)$.

THEOREM 7. *Let $Q$ be a mutation rate matrix that satisfies Assumption 1. Then there exists an $a > 0$, constants $c > 0, \eta > 0$, two trees $T_1, T_2$ and open sets $S_1 \subset (0, \infty)^{E(T_1)}, S_2 \subset (0, \infty)^{E(T_2)}$ such that if $\mathbf{l}_1 \in S_1, \mathbf{l}_2 \in S_2$, then for the distribution $\mathcal{D}_1$ generated at the leaves of $(T_1, \mathbf{l}_1)$ and the distribution $\mathcal{D}_2$ generated at the leaves of $(T_2, \mathbf{l}_2)$, the following holds for $\mathcal{D} = (0.5 - \rho)\mathcal{D}_1 + (0.5 + \rho)\mathcal{D}_2$ for $|\rho| < \eta$ and all $\varepsilon > 0$:*

*Let $\mathbf{D} = (D_1, \ldots, D_N)$, chosen independently from the distribution $\mathcal{D}$. Consider a Markov chain on discrete or continuous tree space with only NNI, SPR or TBR transitions. Then with probability $1 - \exp(-cN)$ over* *the data generated, the mixing time of the Markov chain, for priors which are* $(a, \varepsilon)$-*regular, satisfies*

$$T_{\mathrm{mix}} \geq c\varepsilon \exp(cN).$$

REMARK 8. It is straightforward to check that Theorem 1 is a special case of Theorem 7. This follows by the standard translation between edge lengths and edge probabilities. As mentioned above, the CFN model (as well as Jukes–Cantor and many other models) satisfies Assumption 1.

**2. Proof of the general theorem.** We begin with the definition of the *mixture distribution* $\mathcal{D}$ for general models. Let $\mathcal{D}_1$ be the distribution at the leaves of the evolutionary model defined on the tree $T_1(a, a^2)$ and $\mathcal{D}_2$ be the distribution at the leaves of the evolutionary model defined on the tree $T_2(a, a^2)$. We let $\mathcal{D} = (0.5 - \rho)\mathcal{D}_1 + (0.5 + \rho)\mathcal{D}_2$.

We first expand the distribution $\mathcal{D}$. This is easily done in terms of $C^*$, where $C^*$ is the set of cherries in $T_1 \cup T_2$. We will use the following definition of a cherry:

DEFINITION 9. Let $T$ be a tree. We say that that a pair of leaves $i, j$ is a *cherry* of $T$ if there exists a single edge $e$ of $T$ such that removing $e$ disconnects $i$ and $j$ from the other leaves of $T$. For a tree $T$, we let $C(T)$ denote the set of cherries of $T$.

Note that according to this definition, the "star" tree has no cherries. We clearly have $C^* = C(T_1) \cup C(T_2) = \{(12), (15), (45), (24)\}$.

Our theorem holds for $a$ sufficiently small. Hence, the asymptotic notation in our proofs is in terms of $1/a \to \infty$. Thus, $a = o(-a \log a)$ and $a^2 = o(-a \log a)$, since $-\log a \to \infty$ as $a \to 0$.

It is easy to estimate $\mathcal{D}$ for small $a$. This follows from the following lemma:

LEMMA 10. *For an edge $e$ of length $b$, conditioned on the character at the endpoint of the edge, the probability that the other endpoint obtains the same label is $1 - b + O(b^2)$. The probability that it obtains a different label is $b + O(b^2)$.*

PROOF. Part 3 of Assumption 1, along with the expansion of $\exp(bQ)$, implies that

$$\exp(bQ) = I + bQ + O(b^2).$$

The lemma now follows from the fact that $\sum_{\beta \neq \alpha} Q_{\alpha,\beta} = 1$, which was stated as part 3 of Assumption 1. □

We begin by compiling some simple facts concerning the mixture distribution $\mathcal{D}$. We will use the following notation for characters. By $\alpha$, we denote the character that is constant $\alpha$. We let $F_\varnothing$ denote the set of all constant characters. By $(\alpha, i, \beta)$, we denote the character that is $\alpha$ on all leaves except $i$, where it is $\beta$. The set of all such characters is denoted by $F_i$. By $(\alpha, i, j, \beta)$, we denote the character that is $\beta$ on $i, j$ and $\alpha$ on all other leaves. The set of all such characters is denoted by $F_{i,j}$. We denote by $G$ the set of all other characters.

LEMMA 11. *The mixture distribution $\mathcal{D}$ satisfies the following conditions:*

- *for all $\alpha \in F_\varnothing$,*

(2) $$\mathcal{D}[\alpha] = \pi_\alpha(1 - 2a + O(a^2));$$

- *for all $i$,*

(3) $$\mathcal{D}[F_i] = O(a^2);$$

- *for all $(i,j) \in C^*$,*

(4) $$\mathcal{D}[F_{i,j}] = a/2 + O(a^2);$$

- *for $(i,j) \notin C^*$,*

(5) $$\mathcal{D}[F_{i,j}] = O(a^2);$$

-

(6) $$\mathcal{D}[G] = O(a^2).$$

PROOF. We first claim that it suffices to assume that the mixture is generated from $T_1(a, a^2)$ and $T_2(a, a^2)$ with equal weights ($\rho = 0$). Since $\mathcal{D}[\sigma]$ is continuous in the edge lengths and the mixture probabilities, proving the bounds (2)–(6) for this mixture implies the same bounds for $(T_1, \mathbf{l}_1)$ and $(T_2, \mathbf{l}_2)$, as long as $\mathbf{l}_i(e)$ is close to $a$ for internal edges and close to $a^2$ for terminal edges. Similarly, the same bounds will hold for mixture probabilities $0.5 - \rho$ and $0.5 + \rho$, as long as $\rho$ is sufficiently small.

Hereafter, a terminal edge refers to an edge connected to a leaf. Consider first the probability $D \notin F_\varnothing$. There are two ways in which this can occur: either there is a mutation on exactly one of the internal edges and no mutations on the terminal edges, which occurs with probability $2a(1-a) + O(a^2) = 2a + O(a^2)$, or there is a mutation on a terminal edge and/or both internal edges, these occurring with probability $O(a^2)$ by our choice of edge lengths on $T_1$ and $T_2$. This implies (2).

For $D \in F_i$, there needs to be a mutation on at least one terminal edge, which implies (3). Next, consider a cherry $(i,j) \in C^*$, say $(i,j) = (1,2) \in$ $C(T_1)$. To generate $D \in F_{i,j}$, the dominant term is if we are generating from $T_1$ and have a mutation on the internal edge to the parent of leaves 1 and 2 or a mutation on more than one terminal branch. Alternatively, if we generate from $T_2$, we need a mutation on more than one terminal branch. We thus obtain (4). It is easy to see that in order to generate $D \in F_{i,j}$ where $(i,j) \notin C^*$ or $D \in G$, two mutations are needed. This implies (5) and (6). □

REMARK 12. Note that the argument above for $\alpha \in F_\varnothing$ can be extended to show the following. Let $T$ be any tree and $\mathbf{l}$ be a vector of edge lengths. Then

$$\mathcal{D}[\alpha] = \pi_\alpha \left(1 - \sum_e \mathbf{l}(e)\right) + O\left(\max_e \mathbf{l}(e)^2\right).$$

In particular, if $a$ is sufficiently small and $\mathbf{l}(e) \leq 4a \log(1/a)$ for all $e$, then

$$\mathcal{D}[\alpha] \leq \pi_\alpha$$

for all $\alpha$.

DEFINITION 13. The expected log-likelihood of a tree $T$ with edge lengths $\mathbf{l}$ is defined to be

$$L_\mathcal{D}(T, \mathbf{l}) = \mathbf{E}_{x \in \mathcal{D}} \log \Pr(x|T, \mathbf{l}).$$

Let $L_\mathcal{D}(T, \ell)$ denote the expected log-likelihood of the tree $T$ with all edge lengths $\ell$. We will show that tree $T_1$ with all edge lengths $a$ (including terminal edges) has large likelihood. We denote by $(T_1, a)$ the tree $T_1$ where all edge lengths are exactly $a$.

LEMMA 14. The tree $T_1$ satisfies

$$L_\mathcal{D}(T_1, a) \geq H(\pi) + (1 + o(1))3a \log a$$

and a similar inequality is satisfied by $T_2$, where

$$H(\pi) = \sum_\alpha \pi_\alpha \log \pi_\alpha.$$

PROOF. We prove the result for $T_1$; the proof for $T_2$ is identical. We first consider the sequences in $F_\varnothing$. By (2), the $\mathcal{D}$-probability of the sequence $\alpha$ is

(7) $\qquad \mathcal{D}[\alpha] = \pi_\alpha(1 - 2a + O(a^2)) = \pi_\alpha + o(a \log a),$

while the log-likelihood of $\alpha$ according to $(T_1, a)$ is given by

$$\log \Pr(\alpha|T_1, a) = \log(\pi_\alpha(1 - 2a + O(a^2)))$$
$$= \log(\pi_\alpha) - 2a + O(a^2)$$
$$= \log(\pi_\alpha) + o(a \log a).$$

Thus, the total contribution to $L_\mathcal{D}(T_1, a)$ coming from $F_\varnothing$ is

$$\sum_{\sigma \in F_\varnothing} \mathcal{D}[\sigma] \log(\Pr(\sigma|T_1, a)) = \sum_\alpha \mathcal{D}[\alpha] \log(\Pr(\alpha|T_1, a))$$
(8)
$$= H(\pi) + o(a \log a).$$

All sequences in $F_i$ provide a contribution of $O(a^2)$ up to log corrections, which is also $o(a \log a)$. Similar situations obtain for sequences in $F_{i,j}$ such that $(i, j) \notin C^*$ and for sequences in $G$. Let

$$F_* = \bigcup_{(i,j) \in C^*} F_{i,j}.$$

Then we have

(9)
$$\sum_{\sigma \in A^5 \setminus (F_\varnothing \cup F_*)} \mathcal{D}[\sigma] \log \Pr(\sigma|T_1, a) = o(a \log a).$$

If $(i, j)$ belongs to $C^*$, there are two possibilities: either $(i, j)$ is a cherry of $T_1$ or it is a cherry of $T_2$. First, if $(i, j)$ is a cherry of $T_1$ and $(\alpha, i, j, \beta) \in F_{i,j}$, then

$$\log(\Pr((\alpha, i, j, \beta)|T_1, a)) = (1 + o(1)) \log a.$$

[This follows by considering a single mutation along the internal edge that separates the cherry $(i, j)$ from the rest of the tree.] For $(i, j) \in C^* \setminus C(T_1)$ [i.e., $(i, j) \in C(T_2)$], we have, for all $(\alpha, i, j, \beta) \in F_{i,j}$ that this character occurs if the only mutations are on the pair of terminal edges connected to $i$ and $j$ or, otherwise, if at least three mutations occurred. Hence,

$$\log(\Pr((\alpha, i, j, \beta)|T_1, a)) = (2 + o(1)) \log a.$$

Since $C^*$ contains two cherries from $T_1$ and two from $T_2$, we obtain

$$\sum_{\sigma \in F_*} \mathcal{D}[\sigma] \log \Pr(\sigma|T_1, a)$$

$$= \sum_{\substack{(\alpha, i, j, \beta): \\ (i,j) \in C(T_1)}} \mathcal{D}[(\alpha, i, j, \beta)] \log(\Pr((\alpha, i, j, \beta)|T_1, a))$$

$$+ \sum_{\substack{(\alpha, i, j, \beta) \\ (i,j) \in C(T_2)}} \mathcal{D}[(\alpha, i, j, \beta)] \log(\Pr((\alpha, i, j, \beta)|T_1, a))$$

(10)
$$= (1 + o(1)) \log a \sum_{\substack{(\alpha, i, j, \beta): \\ (i,j) \in C(T_1)}} \mathcal{D}[(\alpha, i, j, \beta)]$$

$$+ (2 + o(1)) \log a \sum_{\substack{(\alpha,i,j,\beta) \\ (i,j) \in C(T_2)}} \mathcal{D}[(\alpha,i,j,\beta)]$$

$$= \left(\frac{a}{2} + O(a^2)\right)(1 + o(1))2(2\log a + \log a)$$

$$= (1 + o(1))3a \log a,$$

where the first inequality in (10) follows from (4). Combining (8), (9) and (10) completes the proof of the lemma. □

REMARK 15. Repeating the proof above shows that

$$L_\mathcal{D}(T_1, \mathbf{l}) \geq H(\pi) + (1 + o(1))3a\log a$$

if all the edge lengths $\mathbf{l}$ are in $[a/2, 2a]$. Note that since $\log a$ is negative,

$$3.1 a \log a < 3a \log a.$$

Hence, for $T = T_1$ or $T = T_2$, if all the edge lengths $\mathbf{l}$ are in $[a/2, 2a]$, for $a$ sufficiently small, we have

(11) $$L_\mathcal{D}(T, \mathbf{l}) \geq H(\pi) + 3.1 a \log a.$$

For tree topologies different from those of $T_1$ and $T_2$, we will bound the maximum of their expected log-likelihood. The analysis considers two cases: either all of the edge lengths are smaller than $O(a \log(1/a))$ or there is at least one long edge. When there is one long edge, we only to consider the contribution from constant characters.

LEMMA 16. *Let $(T, \mathbf{l})$ be a tree such that at least one of the edge lengths is greater than $4a\log(1/a)$. If $a$ is sufficiently small, then the following holds. For all constant characters $\alpha$,*

(12) $$\log(\Pr(\alpha|T,\mathbf{l})) \leq \log(\pi_\alpha(1 + 3.9a\log a))$$

*and the total contribution from all constant characters is*

(13) $$\sum_\alpha \mathcal{D}(\alpha) \log(\Pr(\alpha|T,\mathbf{l})) \leq \sum_\alpha \mathcal{D}(\alpha) \log(\pi_\alpha(1 + 3.9a\log a))$$
$$\leq H(\pi) + 3.8 a \log a.$$

PROOF. Let $T$ be any tree for which the sum of the edge lengths is more than $4a\log(1/a)$. To generate the constant character $\alpha$, either the root chooses $\alpha$ and there are no mutations, or there are at least two mutations. Hence,

$$\Pr(\alpha|T,\mathbf{l}) \leq \pi_\alpha(1 - 4a\log(1/a)) + o(a\log(1/a))$$
$$= \pi_\alpha(1 + 4a\log a) + o(a\log a).$$

Hence, for $a$ sufficiently small, (12) holds.

Recall from (7) that $\mathcal{D}[\alpha] = \pi_\alpha + o(a \log a)$. Now, we have

$$\sum_\alpha \mathcal{D}(\alpha) \log(\Pr(\alpha|T,\mathbf{l})) \leq \sum_\alpha \mathcal{D}(\alpha) \log(\pi_\alpha(1 + 3.9a \log a))$$

$$\leq \sum_\alpha (\pi_\alpha + o(a \log a))(\log \pi_\alpha + 3.9a \log a)$$

$$= \sum_\alpha \pi_\alpha \log(\pi_\alpha) + 3.9a \log a \sum_\alpha \pi_\alpha + o(a \log a)$$

$$= H(\pi) + 3.9a \log a + o(a \log a).$$

For $a$ sufficiently small, (13) follows. □

When we restrict our attention to trees all of whose edge lengths are at most $4a \log(1/a)$, we need to consider the contribution from constant characters and characters where one cherry differs from the rest of the tree.

LEMMA 17. *Let $(T,\mathbf{l})$ be a tree all of whose edge lengths are at most $4a\log(1/a)$ and suppose further that $T$ has a topology different from those of $T_1$ and $T_2$. Then the following hold for sufficiently small $a$:*

- *for all constant characters $\alpha$,*

(14) $$\log(\Pr(\alpha|T,\mathbf{l})) \leq \log(\pi_\alpha);$$

- *for all cherries $(i,j) \in C^* \setminus C(T)$ and all $\alpha \neq \beta$,*

(15) $$\log(\Pr((\alpha,i,j,\beta)|T,\mathbf{l})) \leq 1.99a \log a;$$

- *for all cherries $(i,j) \in C^* \cap C(T)$ and all $\alpha \neq \beta$,*

(16) $$\log(\Pr((\alpha,i,j,\beta)|T,\mathbf{l})) \leq 0.99a \log a;$$

- *finally, the total contribution from all constant characters and characters with one cherry different from the rest (i.e., $F_\varnothing \cup F_*$) is*

(17)
$$\sum_{\sigma \in F_\varnothing \cup F_*} \mathcal{D}(\sigma) \log(\Pr(\sigma|T,\mathbf{l}))$$
$$\leq \sum_\alpha \mathcal{D}(\alpha) \log \pi_\alpha + 1.99a \log a \sum_{\substack{(\alpha,i,j,\beta):\\(i,j) \in C^* \setminus C(T)}} \mathcal{D}[(\alpha,i,j,\beta)]$$
$$+ 0.99a \log a \sum_{\substack{(\alpha,i,j,\beta):\\(i,j) \in C^* \cap C(T)}} \mathcal{D}[(\alpha,i,j,\beta)]$$
$$\leq H(\pi) + 3.45a \log a.$$

Before proving the above lemma, we make the following combinatorial observation:

OBSERVATION 18. *Let $T \neq T_1, T_2$. Then*
$$|C(T) \cap C^*| \leq 1.$$

The observation follows by considering a tree $T$ that contains at least one of the cherries of $C^*$, say $(1,2)$. Clearly, $T$ can not also contain the cherry $(1,5)$ or the cherry $(2,4)$. And if it contains the cherry $(4,5)$, then $T = T_1$.

PROOF OF LEMMA 17. The bound (14) follows from Remark 12.

We will now bound the contribution to $L_{\mathcal{D}}(T, \mathbf{l})$ provided by the cherries in $C^*$. Note that if $(i,j) \in C^*$ and $(i,j) \notin C(T)$, then for all $\alpha$ and $\beta$, to generate $(\alpha, i, j, \beta)$, we need at least two mutations. Hence,

$$\log(\Pr((\alpha, i, j, \beta)|T, \mathbf{l})) \leq \log(O(a^2 \log^2(1/a)))$$
$$= 2(1 + o(1)) \log a$$
$$\leq 1.99 a \log a,$$

for sufficiently small $a$. If $(i,j) \in C(T) \cap C^*$, then we need a mutation on the internal edge to the cherry $(i,j)$ (or we need at least two mutations). Hence,

$$\log(\Pr((\alpha, i, j, \beta)|T, \mathbf{l})) \leq \log(4a \log(1/a)) + o(a \log(1/a))$$
$$= (1 + o(1)) \log a$$
$$\leq 0.99 a \log a,$$

for sufficiently small $a$. Combining these inequalities with (4) and Observation 18, which says that there are at least three cherries in $C^* \setminus C(T)$, we have

$$\sum_{\sigma \in F_\varnothing \cup F_*} \mathcal{D}(\sigma) \log(\Pr(\sigma|T, \mathbf{l}))$$
$$\leq \sum_\alpha \mathcal{D}(\alpha) \log(\Pr(\alpha|T, \mathbf{l}))$$
$$+ \sum_{\substack{(\alpha, i, j, \beta): \\ (i,j) \in C^*}} \mathcal{D}((\alpha, i, j, \beta)) \log(\Pr((\alpha, i, j, \beta)|T, \mathbf{l}))$$
$$\leq \sum_\alpha \mathcal{D}(\alpha) \log \pi_\alpha + 1.99 a \log a \sum_{\substack{(\alpha, i, j, \beta) \\ (i,j) \in C^* \setminus C(T)}} \mathcal{D}[(\alpha, i, j, \beta)]$$

$$+ 0.99a \log a \sum_{\substack{(\alpha,i,j,\beta): \\ (i,j) \in C^* \cap C(T)}} \mathcal{D}[(\alpha,i,j,\beta)]$$

$$\leq H(\pi) + o(a \log a) + (1 + o(1))\frac{a}{2}(3 \times 1.99 \log a + 0.99 \log a)$$

$$\leq H(\pi) + 3.45 a \log a,$$

for sufficiently small $a$. $\square$

DEFINITION 19. Let $\mathbf{D} = (D_1, \ldots, D_N)$ consist of $N$ characters. We let

$$L_{\mathbf{D}}(T, \mathbf{l}) = \sum_{D \in \mathbf{D}} \log(\Pr(D | T, \mathbf{l})).$$

Using the Chernoff bound, together with the previous lemmas, we get the following lemma:

LEMMA 20. *Suppose that $\mathbf{D}$ is drawn according to $N$ independent samples from the distribution $\mathcal{D}$. Then, with probability $1 - e^{-\Omega(N)}$, for all trees $(T, \mathbf{l})$ with the topology of $T_1$ or $T_2$ and edge lengths $\mathbf{l}(e)$ in $[a/2, 2a]$ for all $e$, we have*

$$(18) \qquad L_{\mathbf{D}}(T, \mathbf{l}) \geq (H(\pi) + 3.2 a \log a) N.$$

*Furthermore, for all trees $(T, \mathbf{l})$ with topologies different from those of $T_1$ and $T_2$, we have*

$$(19) \qquad L_{\mathbf{D}}(T, \mathbf{l}) \leq (H(\pi) + 3.3 a \log a) N.$$

PROOF. Note that there is a positive number $\mathcal{D}_{\min}$ such that if all edge lengths of the tree are in $[a/2, 4a \log(1/a)]$ for all $e$, then for each $\sigma \in \mathcal{A}^5$, the probability that $\sigma$ is generated according to $T$ is at least $\mathcal{D}_{\min}$. Moreover, $\mathcal{D}_{\min}$ depends only on $a$ and the matrix $Q$.

Let

$$X(\sigma) = |\{i : D_i = \sigma\}|$$

denote the number of characters generated under the mixture distribution $\mathcal{D}$ whose assignment is $\sigma$. By the Chernoff bound (e.g., [15], Corollary 2.3),

$$\Pr(|X(\sigma) - \mathcal{D}(\sigma)N| \geq \varepsilon \mathcal{D}(\sigma)N) \leq 2 \exp(-\varepsilon^2 \mathcal{D}(\sigma)N/3).$$

Let

$$(20) \qquad \delta < \delta_1 := \frac{0.1 a \log(1/a)}{|\mathcal{A}|^5 \log(1/\mathcal{D}_{\min})}.$$

LIMITATIONS OF PHYLOGENETIC MCMC                    17Taking a union bound over the $|\mathcal{A}|^5$ assignments, we have

$$\Pr(\text{for all } \sigma \in \mathcal{A}^5, |X(\sigma) - \mathcal{D}(\sigma)N| \leq N\delta) \geq 1 - \exp(-\Omega(N)).$$

Now, if $T$ is $T_1$ or $T_2$ and $\mathbf{l}$ is such that $\mathbf{l}(e)$ is in $[a/2, 2a]$ for all $e$, then, with probability $1 - \exp(-\Omega(n))$, we have

$$L_{\mathbf{D}}(T,\mathbf{l}) = \sum_{\sigma \in \mathcal{A}^5} X(\sigma) \log \Pr(\sigma|T,\mathbf{l})$$

$$\geq \sum_{\sigma \in \mathcal{A}^5} N(\mathcal{D}(\sigma) + \delta) \log \Pr(\sigma|T,\mathbf{l})$$

$$= NL_{\mathcal{D}}(T,\mathbf{l}) + \delta N \sum_{\sigma \in \mathcal{A}^5} \log \Pr(\sigma|T,\mathbf{l})$$

$$\geq NL_{\mathcal{D}}(T,\mathbf{l}) + \delta N |\mathcal{A}|^5 \log(\mathcal{D}_{\min})$$

$$\geq (H(\pi) + 3.1a \log a + 0.1a \log a)N$$

$$= (H(\pi) + 3.2a \log a)N,$$

the last inequality following from (11) and (20). This proves (18).

For the proof of (19), we have to consider two cases. The first case is where at least one of the edges of $T$ is of length $> 4a \log(1/a)$. In this case, except with probability $\exp(-\Omega(n))$, we have

$$L_{\mathbf{D}}(T,\mathbf{l}) = \sum_{\sigma \in \mathcal{A}^5} X(\sigma) \log \Pr(\sigma|T,\mathbf{l})$$

$$\leq \sum_{\alpha} X(\alpha) \log(\pi_\alpha(1 + 3.9a \log a)) \qquad \text{by (12)}$$

$$\leq \sum_{\alpha} N(\mathcal{D}(\alpha) - \delta) \log(\pi_\alpha(1 + 3.9a \log a))$$

$$\leq N\left(H(\pi) + 3.8a \log a - |\mathcal{A}|\delta\left(\max_\alpha \log(\pi_\alpha(1 + 3.9a \log a))\right)\right),$$

the last inequality following from (13). Therefore, if we take

$$\delta < \delta_2 := \frac{0.1a \log(1/a)}{|\mathcal{A}| \max_\alpha \log(\pi_\alpha(1 + 3.9a \log a))},$$

then we obtain

$$L_{\mathbf{D}}(T,\mathbf{l}) \leq N(H(\pi) + 3.7a \log a),$$

with probability $1 - \exp(-\Omega(n))$.

The second case is where all of the edge lengths are of length at most $4a \log(1/a)$. We now have

$$L_{\mathbf{D}}(T,\mathbf{l}) = \sum_{\sigma \in \mathcal{A}^5} X(\sigma) \log \Pr(\sigma|T,\mathbf{l})$$

$$\leq \sum_\alpha X(\alpha) \log \Pr(\alpha|T,\mathbf{l}) + \sum_{\substack{\alpha,\beta,i,j:\\ \alpha \neq \beta, i \neq j}} X(\alpha,i,j,\beta) \log \Pr((\alpha,i,j,\beta)|T,\mathbf{l})$$

$$\leq \sum_\alpha X(\alpha) \log \pi_\alpha + 1.99 a \log a \sum_{\substack{(\alpha,i,j,\beta):\\ (i,j) \in C^* \setminus C(T)}} X(\alpha,i,j,\beta)$$

$$+ 0.99 a \log a \sum_{\substack{(\alpha,i,j,\beta):\\ (i,j) \in C^* \cap C(T)}} X(\alpha,i,j,\beta)$$

$$\leq \sum_\alpha N(\mathcal{D}(\alpha) - \delta) \log \pi_\alpha$$

$$+ 1.99 a \log a \sum_{\substack{(\alpha,i,j,\beta):\\ (i,j) \in C^* \setminus C(T)}} N(\mathcal{D}(\alpha,i,j,\beta) - \delta)$$

$$+ 0.99 a \log a \sum_{\substack{(\alpha,i,j,\beta):\\ (i,j) \in C^* \cap C(T)}} N(\mathcal{D}(\alpha,i,j,\beta) - \delta)$$

$$\leq N\Big(H(\pi) + 3.45 a \log a + 20\delta|\mathcal{A}|^2 a \log(1/a) - \delta|\mathcal{A}| \max_\alpha \log(\pi_\alpha)\Big),$$

where the second inequality follows from (14), (15) and (16) and the last inequality follows from (17). Therefore, choosing

$$\delta < \delta_3 := \min\bigg(\frac{0.05}{20|\mathcal{A}|^2}, \frac{0.05 a \log(1/a)}{\max_\alpha - \log(\pi_\alpha)}\bigg),$$

we obtain that, except with probability $\exp(-\Omega(n))$, we have

$$L_\mathbf{D}(T,\mathbf{l}) \leq N(H(\pi) + 3.4 a \log a),$$

as needed. Taking $\delta = \min(\delta_1, \delta_2, \delta_3)$ concludes the proof of the lemma. $\square$

LEMMA 21. *Let $\varepsilon > 0$ and let $\Psi$ be an $(\varepsilon, a)$-regular prior on $\mathcal{T}$. Then, with probability $1 - e^{-\Omega(N)}$, if $T \neq T_1, T_2$,*

$$\frac{w(T)}{w(T_1)} \leq \frac{a^{-7}}{\varepsilon} \exp(-0.1 a \log(1/a) N).$$

PROOF. With probability $1 - e^{-\Omega(N)}$, we have that (18) and (19) hold. Since $\Psi$ is $(\varepsilon, a)$-regular we see that

$$w(T_1) = \int_\mathbf{l} \exp(L_\mathbf{D}(T_1, \mathbf{l})) \Psi(T_1, \mathbf{l}) \, d\mathbf{l}$$

$$\geq \varepsilon a^7 \exp(H(\pi) N) \exp((3.2 a \log a) N).$$

On the other hand,
$$w(T) = \int_{\mathbf{l}} \exp(L_{\mathbf{D}}(T, \mathbf{l})) \Psi(T, \mathbf{l}) \, d\mathbf{l}$$
$$\leq \exp(H(\pi)N) \exp((3.3a \log a)N).$$

The claim follows. □

To complete the proof of Theorem 1, we need the well know fact that small conductance implies slow mixing. This is standard for discrete spaces; see, for example, [5]. Since we consider also the continuous case, we prove the following claim below.

LEMMA 22. *Consider a discrete-time Markov chain $P$ on a discrete or continuous state space with a unique stationary measure $\mu$. Assume, furthermore, that there exists a partition of the state space into three sets $A_1$, $A_2$ and $B$ such that the probability of a move from $A_2$ to $A_1$ is 0 [in the sense that $\int d\mu(x) 1(x \in A_2) \int dP(x,y) 1(y \in A_1) = 0$] and $\mu(A_1) \geq \mu(A_2)$, $\mu(B)/\mu(A_i) \leq \varepsilon$ for $i = 1, 2$.*

*Let $\mu^t$ denote the distribution of the chain after $t$ steps, where the initial distribution $\mu^0$ is given by $\mu$, conditioned to $A_2$. Then the total variation distance between $\mu^{1/3\varepsilon}$ and $\mu$ is at least $1/3$.*

PROOF. Let $t = 1/3\varepsilon$ and consider sequences $(x_1, \ldots, x_t)$ of trajectories of the chain, where $x_1$ is chosen according to the stationary distribution. Since each $x_i$ is distributed according to the stationary distribution, the fraction of sequences that contain an element of $B$ is, by the union bound, at most $t\varepsilon\mu(A_2) = \mu(A_2)/4$. The fraction of sequences that have their first element in $A_2$ is $\mu(A_2)$. Thus, conditioned on having $x_1 \in A_2$, the probability that $x_t \in B \cup A_1$ is at most $1/3$. Since the stationary measure of $B \cup A_1$ is at least $1/2$, the claim follows. □

PROOF OF THEOREM 1. The proof now follows from Lemmas 21 and 22 — we take the two sets corresponding to $T_1$ and $T_2$ with all edge lengths strictly between 0 and $\infty$. The proof follows from the observation that $T_1$ and $T_2$ are not connected by either NNI, SPR or TBR transitions. □

**3. Future directions.** A popular program is MrBayes [13], which additionally uses what is known as Metropolis-coupled Markov chain Monte Carlo, referred to as MC$^3$ [10]. Analysis of this approach requires more detailed results and it is unclear whether our techniques can be extended this far. Some theoretical work analyzing MC$^3$ in a different context was done by Bhatnagar and Randall [1].

An interesting future direction would be to prove a positive result. In particular, is there a class of trees where we can prove fast convergence to the stationary distribution when the data are generated by a tree in this class? More generally, if the data are generated by a single tree, do the Markov chains always converge quickly to their stationary distribution?

**Acknowledgments.** We thank Bernd Sturmfels and Josephine Yu for interesting discussions on the algebraic geometry of tree space. We also thank the referee for many useful comments.

LIMITATIONS OF PHYLOGENETIC MCMC 21[16] LARGET, B. and SIMON, D. L. (1999). Markov chain Monte Carlo algorithms for the Bayesian analysis of phylogenetic trees. *Mol. Biol. Evol.* **16** 750–759.
[17] LI, S., PEARL, D. K. and DOSS, H. (2000). Phylogenetic tree construction using Markov chain Monte Carlo. *J. Amer. Statist. Assoc.* **95** 493–508.
[18] NEI, M. and KUMAR, S. (2000). *Molecular Evolution and Phylogenetics*. Oxford Univ. Press.
[19] NEYMAN, J. (1971). Molecular studies of evolution: A source of novel statistical problems. In *Statistical Decision Theory and Related Topics* (S. S Gupta and J. Yackel, eds.) 1–27. Academic Press, New York. MR0327321
[20] MOSSEL, E. and VIGODA, E. (2005). Phylogenetic MCMC algorithms are misleading on mixtures of trees. *Science* **309** 2207–2209.
[21] RANNALA, B. and YANG, Z. (1996). Probability distribution of molecular evolutionary trees: A new method of phylogenetic inference. *J. Mol. Evol.* **43** 304–311.
[22] SIMON, D. L. and LARGET, B. (2000). Bayesian analysis in molecular biology and evolution (BAMBE). Version 2.03 beta, Dept. Mathematics and Computer Science, Duquesne Univ., Pittsburgh, PA.
[23] SPEYER, D. and STURMFELS, B. (2004). The tropical Grassmannian. *Adv. Geom.* **4** 389–411. MR2071813
[24] YANG, Z. (2000). Complexity of the simplest phylogenetic estimation problem. *Proc. R. Soc. Lond. B Biol. Sci.* **267** 109–116.
[25] YANG, Z. and RANNALA, B. (1997). Bayesian phylogenetic inference using DNA sequences: A Markov chain Monte Carlo method. *Mol. Biol. Evol.* **14** 717–724.DEPARTMENT OF STATISTICS
UNIVERSITY OF CALIFORNIA AT BERKELEY
BERKELEY, CALIFORNIA 94720
USA
E-MAIL: mossel@stat.berkeley.edu

COLLEGE OF COMPUTING
GEORGIA INSTITUTE OF TECHNOLOGY
ATLANTA, GEORGIA 30332
USA
E-MAIL: vigoda@cc.gatech.edu